\documentclass[]{spie}  %>>> use for US letter paper
\pdfoutput=1
\usepackage[]{graphicx}

\usepackage{upgreek}
\usepackage{color}
\usepackage{hyperref}

\newcommand{\pasp}{PASP}

\newcommand{\doi}[1]{\href{http://dx.doi.org/#1}{doi:#1}}
\newcommand{\arxiv}[1]{\href{http://arxiv.org/abs/#1}{arXiv:#1}}

\title{Your data is your dogfood: DevOps in the astronomical observatory}

\author{Frossie~Economou\supit{a},
Joshua~C.~Hoblitt\supit{a},
Pat~Norris\supit{a}
\skiplinehalf
\supit{a}National Optical Astronomy Observatory, 950 N.\ Cherry Ave, Tucson, AZ 85719, USA
}

\authorinfo{Further author information: (Send correspondence to F.E.)\\F.E.: E-mail: frossie@noao.edu}
\begin{document}
  \maketitle

%%%%%%%%%%%%%%%%%%%%%%%%%%%%%%%%%%%%%%%%%%%%%%%%%%%%%%%%%%%%%
\begin{abstract}
  DevOps is the contemporary term for a software development culture
  that purposefully blurs distinction between software development and
  IT operations by treating ``infrastructure as code.'' DevOps teams
  typically implement practices summarised by the colloquial directive
  to ``eat your own dogfood;'' meaning that software tools developed
  by a team should be used internally rather thrown over the fence to
  operations or users. We present a brief overview of how DevOps
  techniques bring proven software engineering practices to IT
  operations. We then discuss the application of these practices to
  astronomical observatories.

\end{abstract}

%>>>> Include a list of keywords after the abstract


\section{A brief Introduction to DevOps concepts}

\subsection{DevOps}

A portmanteau of the words ``Development'' and ``Operations,'' DevOps is
sometimes mistaken as being simply the principle of having software
development and IT operations performed in the same organisational
group. In fact DevOps is now a term often used to describe a
collection of various techniques, tools and processes that
collectively are a set of best practices for modern IT administration.

DevOps is strongly philosophically aligned with the Agile methodology;
in fact the movement was conceived as ``Agile Infrastructure'' or
``Agile Operations'' prior to the coinage of the term DevOps, and like
the Agile methodology, derives from ideas and practices that had been
present in the computing field long before then, but had yet to be
drawn into a self-identified movement.\footnote{The term ``Agile
  Infrastructure'' was extant by 2005\cite{Berteig}, and by 2008 had influenced a
  number of people including Patrick Debois\cite{Debois}, who later
  coined the term DevOps from a 2009 talk by John Allspaw entitled
  ``10+ Deploys Per Day: Dev and Ops Cooperation at
  Flickr.''\cite{History,Allspaw}}

An explicit early goal of DevOps was to address the ``Wall Of
Confusion''\cite{Edwards}, with software ``thrown over the wall'' from
development to operations. Differing toolsets used by parties on
either side of this wall result in slow and/or fragile deployments,
and initially DevOps advocated unifying systems, processes and often
people in order to bring the deployment and management under the same
agile cyclical process as software development.

Partly as a consequence of developing a number of software and
techniques in pursuing that goal, and partly by other aligned
interests of DevOps practitioners, DevOps is now commonly understood
to encompass a number of related technical goals under one umbrella of
practice.

\subsection{Infrastructure as code}

The explosive growth in computing infrastructure and the demand for
scalability added another dimension to the DevOps approach. The
traditional methods of system administration that revolved around
dealing with individual machines gave way to automated configuration
and deployment systems designed to deal with large numbers of machines
at scale, without an administrator having to ever log onto an
individual machine. Automated configuration systems such as
Puppet\footnote{\url{http://puppetlabs.com}} borrow from
object-oriented patterns and treat computing resources similar to
objects in a class; a machine is no longer a database server, but an
instance of a database server class, with its ultimate configuration
matching a state described in source code and addressed with normal
code-management practices.

The advantages of such a ``codification'' of IT infrastructure are
compelling and include:

\subsubsection{Scalability}

Once a machine's desired end state has been captured in an automated
configuration system, spinning up 5 or 50 of them becomes merely a
matter of procuring hardware resources. The effort associate with
either scenario is the same.

\subsubsection{Version control}

DevOps practitioners typically favour feature-rich version control
systems, the most popular being
\texttt{git}\footnote{\url{http://git-scm.com}}.  The ``Infrastructure
as Code'' approach results in IT configurations being manageable with
the same kind of process as traditional software development such as
release management via repository branching, the ability to version
environments, and easily revert changes. The same benefits accrue:
higher productivity with lower technical risk (high-performing DevOps
organisations report deploying code 30 times more frequently than their
peers\cite{StateDevops13}).

\subsubsection{Agile code quality practices}

Other common practices aligned with Agile methodologies, such as unit
testing, refactoring, and code review/retrospectives, can be brought
to bear to the codified IT infrastructure, bringing many of the same
advantages of robustness and versatility (high-performing DevOps
organisations report double the change success rate of their
peers\cite{StateDevops13}).

\subsubsection{Encapsulation}

By pursuing deployment on environments distinct from bare metal,
DevOps practitioners have embraced a number of deployment technologies
such as virtual machine hypervisors and application containers. Such
techniques allow for a more controlled and secure runtime environment.

\subsubsection{Disaster recovery}

In a DevOps environment the organisation's IT structure can be
replicated on new hardware from just the code repository and
application data backups. This is cheaper, faster and more
reproducible than restoring from system-level backups. Moreover
transitioning a service to different hardware is a simple operation
(high-performing DevOps organisations report 12 times faster service
recovery than their peers\cite{StateDevops13}).

\subsubsection{Efficiency}

The tools in the DevOps arsenal are resulting in unprecedented
machines:staff ratio efficiencies. As recently as 2002, IT
publications carried articles that said:\cite{zdnet}

\begin{quote}
Our research indicates that most IT organizations have a
system-to-sys admin range of between 10:1 and 20:1. Accordingly, we
believe shops with a server-to-sys admin ratio greater than 30:1 can
make a good case for additional sys admin staff.
\end{quote}

For comparison purposes, Facebook was reporting in 2013 ratios
exceeding 20,000:1.\cite{Facebook}

\subsubsection{Elasticity of Demand}

The emergence of open-source cloud computing platforms such as
OpenStack\footnote{\url{http://www.openstack.org}}, with compatible
APIs as commercial web services, allow for a heretofore
unprecedented ability to respond elastically to demand; instead of
over-provisioning in-house hardware resources to deal with worst-case
demand scenarios whilst having them sit idle for periods of trough
demand, application instances can be spun up on a variety of
additional platforms to service peak demand.

\subsubsection{Team flexibility}

By converging the skills required for software development and system
administration, the team gains flexibility in shifting its focus from
one area to the other, as project demands fluctuate.

\subsubsection{Documentation as Code}

IT systems are changeable (and if not, they are ossified, which bears
its own risks) and notoriously poorly documented. One of the
advantages in capturing machine state in configuration management
systems is that the code becomes the ``documentation'' for that system
--- a reference, guaranteed to be correct, to how the machine or
service is set up.

\subsection{Eating your own dogfood}

DevOps teams typically implement practices summarised by the
colloquial directive urging one to ``eat your own dogfood;'' meaning
that software chains developed by a team should be used internally
rather than with the only intention to be thrown over the fence to
operations, users or customer.

The reason ``dogfooding'' is a major underpinning of DevOps principles
lies both in the DevOps goal of avoiding the ``Wall of Confusion'' as
well as its Agile-inspired roots. When a development team consumes its
own product, it is far more likely to encounter flaws prior to the
release to an operations team. Similarly, the use of the product
within the team allows for the Agile approaches towards it; instead of
the goal being ``delivery,'' internal use promotes Agile virtues such as
introspection, Kanban-like addressing of pain points, and a positive
attitude towards refactoring.

From the technical management point of view, developing directly on the
production technology stack reduces technical risk, speeds up
deployments, increases sources of technical support, and makes the
cost of change cheaper -- the latter being a critical component of the
Agile ethos.

In a way, ``dogfooding'' results in a development team becoming a
stakeholder in its own product.

\subsection{Monitoring}

With roots in Agile philosophies that value continuous improvement and
elimination of bottlenecks, it is unsurprising that DevOps embraces
the generation and monitoring of metrics and associated tools such as
log collectors (e.g., \texttt{logstash}\footnote{\url{http://logstash.net}}),
statistics collectors (e.g., \texttt{statsd}\footnote{\url{https://github.com/etsy/statsd/}}),
log analyzers (e.g., \texttt{kibana}\footnote{\url{http://www.elasticsearch.org/overview/kibana/}}),
log searching (e.g., \texttt{elasticsearch}\footnote{\url{http://www.elasticsearch.org}})
and dashboard-style visualisation (e.g.,
\texttt{graphite}\footnote{\url{http://graphite.readthedocs.org}}).

Goals associated with infrastructure monitoring include:

\begin{itemize}
\item increasing service availability
\item aiding fault diagnosis
\item identifying performance bottlenecks
\item improving system architecture
\end{itemize}

In a DevOps environment, where software development occurs with the
same technology stack as production, these type of metrics can be used
to anticipate and fix problems prior to release into production.

\subsection{The One Machine Conundrum}

An interesting illustration of how DevOps has ``down-evolved'' from a
set of techniques for large-scale IT systems into a set of toolsets
and processes that represent best practice lies in the question:
``Suppose you just had a single machine. Would you go all-DevOps on
it?''

The answer from DevOps practitioners is frequently ``yes,'' and it is
worth briefly exploring why this is.

There is an erroneous perception that operating a system under DevOps
practices is ``more work'' than old-school system administration, and
that these techniques only start paying off at scale. In fact for a
practitioner familiar with the tools of the trade, the work of setting
up one machine is not in any way excessive, since they would almost
certainly already have all the necessary code in their repositories,
or could utilise the extensive public repositories of contributed
software for common unix administration tasks. More to the point, the
effort is simply front-loaded; managing a machine with the DevOps
toolset may involve a higher up-front cost, but greatly reduces the
downstream cost of managing that machine or service.

From the engineer's point of view, this allows them to reduce the
``background noise'' of ongoing machine administration, as many of the
functions demanding ongoing attention with traditional system
administration practices are tedious, can lead to system fragility,
and tend to displace more challenging and rewarding system
improvements that increase job satisfaction.\cite{StateDevops}. In a
typical system under configuration management, third-party modules are
frequently updated by their contributors to perform any necessary
changes in managed packages, further reducing the on-going support effort. 

From a project management point of view this is also a highly
desireable approach; if one was to see ongoing support of deployed
services as a kind of technical debt, front-loading IT activities
allows the majority of the effort for administrating that service to
be planned, expended, and accounted for in a more constrained period
of time, thus reducing downstream effort requirement uncertainties.

It is therefore not the case that DevOps techniques are appropriate
only in situations were a single service needs to be scaled, but
rather that they provide a way of engineering IT infrastructure, no
matter how small or diverse in a way that makes it better suited to
software development projects.

\section{DevOps and Astronomy}

\subsection{DevOps and Data Management in Astronomy}

The advent of large-scale data science and the evolution of astronomy
towards data-intensive experiments such as the Large Synoptic Survey
Telescope\cite{LSST,2012SPIE.8451E..0VF}, have brought astronomy much
closer to the technical space occupied by dotcoms as well as other
scientific IT-intensive areas such as particle physics and biology.

On the one hand, it is worth noting the enormous benefits of such
convergence, in particular our ability to utilise a great number of
well supported, rapidly evolving tools and platforms that greatly
increase our ability to manage a data center with the kind of staffing
that academic funding allows. For example, at NOAO's Science Data
Management group, we operate a service that includes 0.7~petabytes of
GPFS-based storage, a processing cluster, a number of
user-facing web and archive services, and our own internal network
architecture, on over 100 machines and VMs, with predominantly a single
DevOps engineer. The level of technical service that can be provided
with modest (though highly-skilled) staffing when leveraging these
technologies is unprecedented.

On the other hand, it means that as a field we are operating in the
same space as commercial entities able to offer much higher
renumeration packages in order to mitigate a skill shortage. It
behooves us to create project and organisational environments that are
attractive to the highly skilled individuals that we will be needing
for our IT activities over the next decade.\cite{ADASS24}

It is worth noting that many of us working in astronomical data
management are undertaking a number of technically ambitious
projects. These projects have IT-centric components, be they software
development, observatory operations or archive services, are critical
to the success of the scientific goals of the project, and the science
value that is returned for public investment. Paying due care to
running our IT infrastructures within the scope of proven best
practices, and aspiring to the technical drive and efficiency that is
characteristic of the more successful young technology companies,
would be a significant contribution to the success of the projects we
serve.

Our users are, after all, operating in the same technological
environment; it will become increasingly difficult to convince an
astronomer that they have to wait a day for an IT service that they
could help themselves to in a few minutes with access to a cloud
service provider and a credit card. As our users become more
accustomed to self-servicing their IT needs, it becomes more important
that archive centers step up to that kind of level of service, or risk
being sidelined by our more motivated users.

\subsection{DevOps and the Observatory as a whole}

While DevOps as an umbrella of practices focuses around software
development and IT infrastructure, as a philosophy it has a relevance
to the kind of wider range of technical and scientific work that takes
place in modern astronomical observatories.

Our observatories are often characterised by functional silos, not
only at the science, hardware, software level (a division that is to
the authors' knowledge ubiquitous), but more problematically within
the software level (telescope, instrument, data reduction, archiving).

Jez Humble writes\cite{Humble}:

\begin{quote}
The DevOps movement addresses the dysfunction that
results from organizations composed of functional silos. [\ldots] DevOps
proposes instead strategies to create better collaboration between
functional silos, or doing away with the functional silos altogether
and creating cross-functional teams (or some combination of these
approaches).
\end{quote}

Unsurprisingly, we see examples of the ``Wall of Confusion'' with every
one of those silos: The instrument throws the data over the wall to
the data reduction which throws the data over the wall to archiving
which throws the data over the wall to the user. If the telescope
pointing is poor, it has to be corrected in data reduction; if the
observing system cannot reliably capture data ownership information,
this has to be fixed by archiving; and so on. Whenever a problem has
to be addressed in a different functional silo than the one
in which it originated, the fix becomes less accurate and far more expensive.

We could benefit from applying some of the DevOps approaches to
dealing with these walls: unified cross-functional
groups,\cite{Pirenne} common toolsets and platforms that allow people
to self-serve their needs, metrics that identify improvements that are
fed back into the groups that can do something about them. Indeed, one
can argue that where many observatories are over-committed or
under-funded, the lean operations model facilitated by
cross-functional, wide-skilled software teams becomes compelling. 

\subsection{Why our data is our dogfood}

The one thing that practically everyone in an observatory ``touches,'' in
one fashion or another, is the data. So if we look to eat our own
dogfood, we have to look at how we can increase internal ``consumption''
of our data products.

There are a number of ways we could use data to erode the boundaries
between our functional silos. Some of them are:

\begin{itemize}
\item Engineering functions are usually poorly served by the scientific
  data flow. In fact there is little reason not to include more
  metadata of engineering interest in data files, and use
  monitoring tools to provide dashboards of quantities of interest
  (e.g., image quality) that are derived in data processing.

\item There is metadata that is of interest to the science user that
  can be captured in the data too, such as weather information, or at
  least associated and published with it in a robust fashion. 

\item It is possible for the archive to collect not only data, but
  time-based information that is of interest to the science user.  For
  example, the information that there was an engineering problem that
  may have affected data quality on the night the dataset they are
  interested in was taken.

\item Create richer, self-defined data formats\cite{Folk:2011:OHT:1966895.1966900,2014Jenness,P91_adassxxiii} and reference libraries to
  access them in order to make data analysis tools useful to a wider
  range of technical and scientific staff.

\item Use DevOps tools and platforms so that observatory or science
  collaboration software can be easily deployed at archive centers,
  user institutions, or the cloud.
\end{itemize}

Treating the data flow as a link between all stakeholders of the
observatory, almost like an API between the telescope/instrument and
the astronomer, becomes even more important as a significant fraction
of science is being done from public data accessed from archives by
astronomers with little knowledge of the originating
facilities.\cite{2012PASP..124..391R,jsahistory}

Perhaps developments in such a direction will also be a positive
direction towards the greatest silos of all: different
observatories. Our field has long bemoaned the low rate of software
re-use in
astronomy,\cite{1998ASPC..145..142M,2002SPIE.4844..321E,2013ASPC..475..383A}
due to a constant re-invention of the wheel, which has held us up as a
discipline. Perhaps as we standardise on current development
techniques, lower barriers to software changes and adopt flexible deployment
practices, we can work as a community towards the kind of software
sharing and continuous progress that we have witnessed in the open
source community. Initiatives such as the Astrophysics Source Code
Library\cite{2014ASPC..485..477A} and moves by observatories, such as
Pan-STARRS\cite{2004AN....325..636H,2007AAS...210.7605P}, LSST\cite{2010SPIE.7740E..38A} and CCAT\cite{jenness_spie2014}, to
open-source and publicly host their software development from the
beginning, are steps in the right direction.

\acknowledgments

FE would like to thank Matt McLeod and Russ Allbery for useful
discussions, and Tim Jenness for assiduous bibliographical services.

%%%%%%%%%%%%%%%%%%%%%%%%%%%%%%%%%%%%%%%%%%%%%%%%%%%%%%%%%%%%%
%%%%% References %%%%%

\end{document}